\documentclass[preprint,showpacs,preprintnumbers,amsmath,amssymb]{revtex4}
\usepackage{graphicx}

\begin{document}
\def \ee {\varepsilon}
\thispagestyle{empty}
{\bf
Comment on ``Contribution of drifting carriers to the Casimir-Lifshitz
and Casimir-Polder interactions with semiconductor materials''
}

It has been shown that the application of the Drude model
in the Lifshitz theory
 leads to problems \cite{1}. The Letter \cite{2}
 modifies the reflection coefficients  by including screening effects.
The modified coefficients  were obtained through use of the Boltzmann
transport equation which takes into account the drift
and diffusion currents.
Here we demonstrate that the inclusion of irreversible
diffusion processes leads to thermodynamic
and experimentally inconsistent results.

The  authors apply their approach  to only
intrinsic semiconductors.
 This approach with the Debye-H\"{u}ckel
screening length $R_{\rm DH}$ is applicable, however, in
all cases where charge carriers of density $n$ are described by
Maxwell-Boltzmann statistics (i.e.,
also for doped semiconductors with $n<n_{cr}$,
some semimetals, and
dielectrics with ionic conductivity). For the latter media
$n$ does not
vanish as $T\to 0$, while the conductivity goes to zero due to the
vanishing mobility. Then, repeating the calculations of \cite{3}
using the approach \cite{2}, one finds a nonzero value of the
 entropy at $T=0$, which depends on the separation, i.e.,
a violation of Nernst's theorem.

In Fig.~1(a) we present the mean measured dif\-fe\-ren\-ces of the
Casimir force between an Au sphere and a Si plate in the presence and
in the absence of light on it \cite{4}.
Computations are done using
\cite{2} and with Si conductivity neglected
in the dark phase \cite{4} (the bands between the dashed and
solid lines, respectively, caused by the uncertainty in $n$).
Both the experimental and theoretical errors
are determined at a 70\% confidence level. At this
confidence the approach of \cite{2} is experimentally excluded.

 The modified reflection coefficients
of \cite{2} are also applicable to metallic plates if $R_{\rm DH}$ is
replaced with the Thomas-Fermi screening length.
In this case, the approach \cite{2} is similar to the Drude model
approach and leads to the same
negative entropy at $T=0$ for perfect crystal lattices
\cite{5}. In Fig.~1(b) we plot
as dots the differences of the theoretical pressures
\cite{2} and the experimental
mean pressures \cite{6} between two
Au plates. It is seen that the
approach \cite{2} is excluded at a 99.9\% confidence level.
The reason for its failure is the inclusion of irreversible processes
 into the Lifshitz formula derived
under the condition of thermal equilibrium.
Drift and diffusion currents are initiated by only {\it external} electric
fields. They lead to unidirectional fluxes of heat from the system to the
heat bath, whereas fluctuating fields lead to equal and mutual exchange
of heat.
\hfill \\[1mm]
\noindent
R.~S.~Decca,${}^1$ E.~Fischbach,${}^2$ B.~Geyer,${}^3$
G.~L.~Klimchitskaya,${}^3$ D.~E.~Krause,${}^{4,2}$ D.~L\'{o}pez,${}^5$
U.~Mohideen,${}^6$
and V.~M.~Mostepanenko${}^3$\hfill \\
${}^1$Department of Physics, Indiana University-Purdue
University Indianapolis, IN 46202, USA\hfill\\
${}^2$Department of Physics, Purdue University, West Lafayette, IN
47907, USA\hfill\\
${}^3$Institute for Theoretical
Physics, Leipzig University,
D-04009, Leipzig, Germany \hfill\\
${}^4$Physics Department, Wabash College, Crawfordsville, IN 47933,
USA\hfill\\
${}^5$Argonne National Laboratory,
Argonne, IL 60439, USA \hfill\\
${}^6$Department of Physics and Astronomy, University of California,
Riverside, CA 92521, USA \hfill

\begin{figure}
\vspace*{-6cm}
\centerline{
\includegraphics{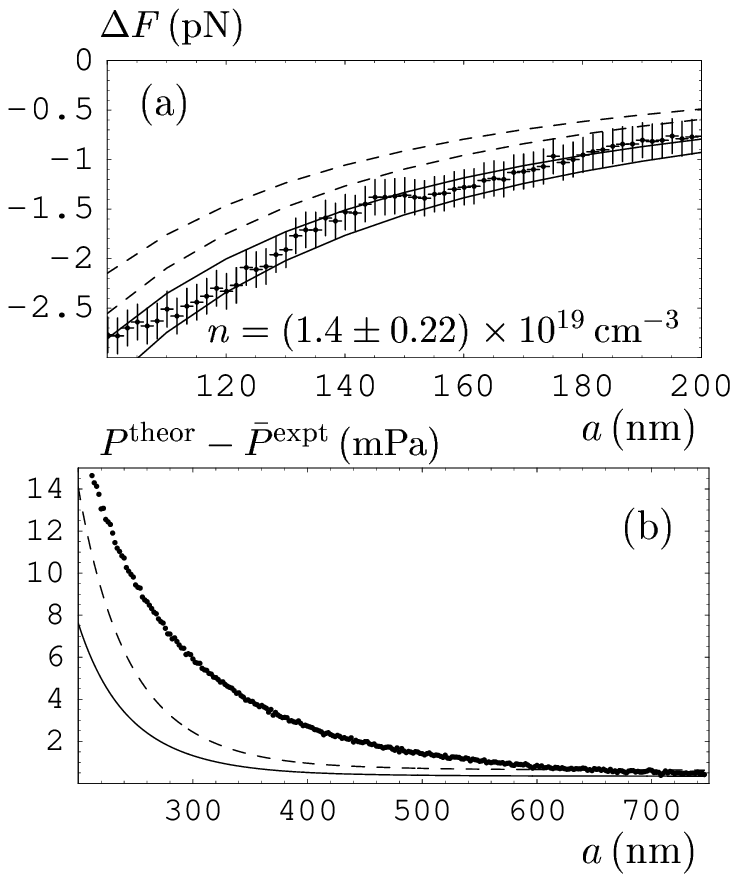}
}
\vspace*{-14cm}
\caption{(a) The measured force differences shown as crosses
versus separation $a$.
The two theoretical bands lie between the solid and dashed lines.
(b) The pressure differences are shown as dots;
the solid and dashed lines indicate confidence intervals at 95\%
and 99.9\%, respectively.}
\end{figure}

\begin{thebibliography}{99}
\bibitem{1}
B.~Geyer {\it et al.},
Phys. Rev. D {\bf 72}, 085009 (2005).
\bibitem{2}
D.~A.~R.~Dalvit and S.~K.~Lamoreaux,
Phys. Rev. Lett. {\bf 101}, 163203 (2008).
\bibitem{3}
G.~L.~Klimchitskaya {\it et al.},
J. Phys. A {\bf 41}, 432001 (2008).
\bibitem{4}
F.~Chen {\it et al.},
Phys. Rev. B {\bf 76}, 035338 (2007).
\bibitem{5}
V. B. Bezerra {\it et al.},
Phys. Rev. A {\bf 69}, 022119 (2004).
\bibitem{6}
R.~S.~Decca {\it et al.},
Phys. Rev. D {\bf 75}, 077101 (2007).
\end{thebibliography}
\end{document}